\documentclass[pra,twocolumn,showpacs]{revtex4-1}
\usepackage{amsmath,amscd,amssymb,color}
\usepackage{graphicx,amsfonts,epsf}
\usepackage{epstopdf}
\usepackage[utf8]{inputenc}
\usepackage[normalem]{ulem}
\usepackage{xcolor}
\usepackage{mathtools}
\usepackage{afterpage}
\usepackage{array}
\usepackage[T1]{fontenc}
\usepackage{float}
\usepackage{hyperref}
\usepackage{multirow}
\usepackage{enumerate}
\usepackage{braket}

\begin{document}
\title{Simulating Three-Flavor Neutrino Oscillations on an NMR
Quantum Processor}
\author{Gayatri Singh}
\email{ph20015@iisermohali.ac.in}
\affiliation{Department of Physical Sciences, Indian
Institute of Science Education \& 
Research Mohali, Sector 81 SAS Nagar, 
Manauli PO 140306 Punjab India.}
\author{Arvind}
\email{arvind@iisermohali.ac.in}
\affiliation{Department of Physical Sciences, Indian
Institute of Science Education \& 
Research Mohali, Sector 81 SAS Nagar, 
Manauli PO 140306 Punjab India.}
\author{Kavita Dorai}
\email{kavita@iisermohali.ac.in}
\affiliation{Department of Physical Sciences, Indian
Institute of Science Education \& 
Research Mohali, Sector 81 SAS Nagar, 
Manauli PO 140306 Punjab India.}
\begin{abstract}
Neutrino oscillations can be efficiently simulated on a quantum computer using
the Pontecorvo-Maki-Nakagawa-Sakata (PMNS) theory in close analogy to
the physical processes realized in experiments. We simulate
three-flavor neutrino oscillations on a two-qubit NMR quantum
information processor. The three-flavor neutrino states were encoded
into the two-qubit system, leaving one redundant basis state
(representing an unphysical sterile neutrino). We simulated the
neutrino oscillations in different scenarios, including propagation in
vacuum and through surrounding matter, and both with and without a CP
violating phase $\delta$ at a Deep Underground Neutrino
Experiment (DUNE) baseline distance of $L=1285$ km.  The oscillation
probabilities were obtained after unitarily time evolving the initial
flavor state and and comparisons were performed between different
scenarios. Further, we design and implement an optimized quantum
circuit encoding all four parameters (three mixing angles-
$\theta_{12},\theta_{23},\theta_{13}$ and one complex phase $\delta$)
to implement the PMNS unitary matrix, which relates the flavor and mass
eigenstates. The interaction with matter is considered as a
perturbation to the vacuum Hamiltonian, and the same quantum circuit is
employed with an approximation of two mixing angles ($\theta_{12}$ and
$\theta_{13}$) and mass eigenvalues. Our experimental results match
well with numerical simulations and highlight the potential of quantum
computers for exploring the physics of neutrino oscillations.
\end{abstract} 
\maketitle 
\section{Introduction}
Quantum computers have the potential to outperform their classical counterparts
through exponential speed-up~\cite{feynman-ijtp-1982} and have emerged as
powerful tools to address complex problems that are either computationally
intractable or challenging to probe experimentally. Quantum
simulators~\cite{llyod-science-1996} can validate theoretical models and
predictions of system behavior in inaccessible physical regimes, which makes
them a promising approach to investigate problems that remain beyond the reach
of current experimental techniques. The increasing interest in quantum
simulation is driven by its broad applicability across numerous fields,
including high-energy physics~\cite{bauer-prxq-2023}, quantum 
chemistry~\cite{kassal-arpc-2011}, cosmology~\cite{mocz-apj-2021}, 
and condensed matter physics~\cite{peng-pra-2005,kazmina-pra-2024}.
Recently, quantum simulations have been used to investigate particle physics,
covering diverse applications such as heavy-ion collision~\cite{jong-prd-2021}, 
neutral Kaon~\cite{bertlmann-pla-2001},
Parton distribution inside proton~\cite{salinas-prd-2021}, 
and neutrino-nucleus
scattering~\cite{roggero-prd-2020}.

The intriguing discovery of neutrino oscillations is a significant breakthrough
in elementary particle physics, and points to the inadequacy of the standard
model (SM)~\cite{cheng-book-1985} and paves the way for exploring physics
beyond the SM~\cite{rasmussen-prd-2017}. Weakly interacting, neutrinos emerged
as a leading candidate for dark matter~\cite{bergstrom-njp-2009} and also play a
pivotal role in astrophysical phenomena such as the merging of neutron star
binaries~\cite{radice-aps-2018} and core-collapse
supernovae~\cite{fuller-prd-1999}. A recent study has demonstrated that neutrino
oscillations violate the Leggett-Garg inequality~\cite{formaggio-prl-2016},
emphasizing their inherent quantum mechanical nature. The idea of the
`neutrino' dates back to W. Pauli~\cite{pauli-book-1991}, who introduced it to
explain the continuous spectrum observed in beta decay. Later, B. Pontecorvo
proposed the idea of neutrino oscillations~\cite{pontecorvo-zetf-1957}, drawing
an analogy to neutral Kaon oscillations, eventually culminating in the theory
of neutrino oscillations in vacuum~\cite{pontecorvo-zetf-1967}. Later, Maki,
Nakagawa, and Sakata proposed the basis misalignment between a neutrino’s
flavor and mass eigenstates in 1962~\cite{maki-ptp-1962}. The propagation of
neutrinos in matter was identified by Wolfenstein~\cite{wolfenstein-prd-1978},
which is now referred to as the `matter effect' and its physical significance
and relevance was provided by Mikheyev and Smirnov~\cite{mikheev-yf-1985}. The
evolution of the neutrino flavor state is described using the
Pontecorvo-Maki-Nakagawa-Sakata (PMNS) matrix, parameterized by three mixing
angles ($\theta_{12},\theta_{13}$ and $\theta_{23},$) and one complex phase
($\delta$). Additionally, the mass squared differences ($\Delta m^2_{ij} =
m^2_i - m^2_j$) are the other three parameters governing neutrino oscillations,
where $m_i$ represents the mass of the $i$th neutrino mass eigenstate.

A hybrid algorithm was employed to simulate the time evolution of collective
neutrino oscillations, with the time-dependent Hamiltonian being approximated
using the Trotter-Suzuki method~\cite{siwach-prd-2023}.  The efficiency of
various quantum hardwares in simulating collective neutrino oscillations was
demonstrated on an IBM quantum
computer~\cite{arguelles-prr-2019,molewski-prd-2022,hall-prd-2021,yeter-qip-2022},
trapped ions~\cite{noh-njp-2012,amitrano-prd-2023,illa-prl-2023} and via
quantum walk frameworks~\cite{molfetta-njp-2016,mallick-epjc-2017}.

The work presented here is the first experimental simulation of neutrino
oscillations on an NMR quantum information processor. We proposed an optimized
quantum circuit, which was decomposed into CNOT gates 
and single-qubit rotations, to
implement the PMNS matrix that encodes all four parameters
$(\theta_{12},\theta_{13},\theta_{23},\delta)$ and implemented it
experimentally. For all the computations and experiments, we adopted all the
parameters as provided in~\cite{esteban-jhep-2020}, including $\Delta
m^2_{21}=7.42 \times 10^{-5} \text{eV}^2$, $\Delta m^2_{31}=2.510 \times
10^{-3} \text{eV}^2$, $\theta_{12}=33.45^\circ$, $\theta_{13}=8.62^\circ$ and
$\theta_{23}=42.1^\circ$. We also experimentally simulated the neutrino
oscillations with a fixed energy ($E=0.5$ GeV) in the presence of matter,
considering potentials $V=5\times 10^{-5}$ eV and $V=10^{-4}$ eV, and compared
the results with the vacuum case ($V=0$ eV). Additionally, the impact of the
CP-violating phase $\delta$ was simulated for the long baseline Deep
Neutrino Underground Experiment (DUNE)~\cite{acciarri-arxiv-2016} ($L=1285$
km) by varying $E$. The results were compared for the vacuum and the
matter-interacting scenarios.

Neutrino oscillations occur because of the mismatch
between the flavor eigen basis and the mass eigen basis in
a three-dimensional Hilbert space, very similar to the
case of quarks. PMNS matrices in this case and the CKM
matrices in the case of quarks, are special matrices
belonging to the group $SU(3)$~\cite{new1,new2}. The study of three-level
systems  and the structure of $SU(3)$ has several interesting
aspects and we will extend our work to explore some of
these directions~\cite{new3,new4,jha-epj-2024}.

This paper is organized as follows: the PMNS theory of three-flavor neutrino
oscillations is briefly described in Section~\ref{sec2}, offering a
mathematical description of neutrino oscillations in vacuum in
Section~\ref{sec2a}. Section~\ref{sec2b} outlines the effective Hamiltonian,
evolution, mixing angle, and mass eigenvalues for neutrinos interacting with
matter. The encoding of three neutrino flavors in a two-qubit system and the
quantum circuit for simulating three-flavor neutrino oscillations is described
in Section~\ref{sec2c}. Section~\ref{sec3a} contains details of the NMR
experiments involving initial state preparation, while Section~\ref{sec3b}
describes the simulation of neutrino oscillations encompassing experimental
results and analysis for all scenarios, both with and without the CP-violating
phase: $(i)$ in vacuum and $(ii)$ interaction with matter. A few conclusions
are presented in Section~\ref{sec4}.

\section{Neutrino Oscillations and PMNS theory}
\label{sec2}
Lepton mixing is a consequence of  the non-correspondence
between the basis of the flavor eigenstate $\{\vert
\nu_e\rangle ,\vert \nu_\mu\rangle, \vert \nu_\tau\rangle\}$
and the basis of the mass eigenstate $\{\vert \nu_1\rangle, \vert
\nu_2\rangle, \vert \nu_3\rangle \}$ for
neutrinos~\cite{giunti-2007,giganti-ppnp-2018}. The
existence of this lepton flavor mixing and non-degenerate
neutrino masses~\citep{bilenky-pr-1978,pontecorvo-zetf-1967}
gives rise to the interesting quantum mechanical phenomenon
known as ``Neutrino oscillations''.  The $SU(3)$ matrix that
connects these two sets of bases and thus accounts for
misalignment between the mass of a neutrino and its flavor
eigenstates is known  as the
Pontecorvo-Maki-Nakagawa-Sakata (PMNS) matrix
($U_{\text{PMNS}}$)~\cite{maki-ptp-1962,giganti-ppnp-2018}.

A
neutrino with flavor $\alpha$ and momentum $p$ is described
using the PMNS matrix in terms of the mass eigen states as
\begin{equation}
\vert \nu_\alpha\rangle=U^*_{\alpha j}\vert\nu_j\rangle
\label{eq1}
\end{equation}
where, $\alpha=e,\mu,\tau$ and $i=1,2,3$, and $U^*_{\alpha
j}$ is $\alpha j$th element of complex conjugate of
$U_{\text{PMNS}}$ matrix (For antineutrinos, we have: $\vert
\bar{\nu}_\alpha\rangle=U_{\alpha
j}\vert\bar{\nu}_j\rangle$).  For Dirac neutrinos, the PMNS
matrix is completely characterized by three mixing angles
$\theta_{12},\theta_{13},\theta_{23}$ and a complex phase
$\delta$ for charge conjugation and parity reversal (CP)
symmetry violations as:
\begin{equation}
\begin{split}
U_{\text{PMNS}}
=\begin{pmatrix}
 U_{e1}&  U_{e2}&U_{e3}  \\
 U_{\mu 1}&  U_{\mu 2}&U_{\mu 3}  \\
 U_{\tau 1}&  U_{\tau 2}&U_{\tau 3}  \\
\end{pmatrix}\equiv U_{\text{PMNS}}(\theta_{12},\theta_{23},\theta_{13},\delta)\\
=\begin{pmatrix}
     1&0&0\\
     0&\cos \theta_{23}&\sin\theta_{23}\\
     0&-\sin \theta_{23}&\cos\theta_{23}\\
     \end{pmatrix}
     \begin{pmatrix}
     \cos \theta_{13}&0&\sin\theta_{13} e^{-\iota\delta}\\
     0&1&0\\
     -\sin \theta_{13} e^{\iota\delta}&0&\cos\theta_{13}\\
     \end{pmatrix}\\
     \begin{pmatrix}
     \cos \theta_{12}&\sin\theta_{12}&0\\
     -\sin \theta_{12}&\cos\theta_{12}&0\\
      0&0&1\\
     \end{pmatrix}   
     \end{split}  \label{eq2}
\end{equation}
\subsection{Neutrino Oscillations in Vacuum}
\label{sec2a}
Since neutrinos are relativistic particles, the energy of the $j$-th mass
eigenstate with momentum $p$ and mass $m_j$ is approximated in vacuum as
$E_j=p c+\frac{m^2_j c^4}{2E}$.
In the PMNS framework, in order to evaluate the time
evolution of an  initial flavor eigen state$\vert
\vert\nu_\alpha (0)\rangle$  (at $t=0$),  we will have to
first expand it in the mass basis, time evolve each
mass eigen state and then re-express the result in the
flavor basis as:
\begin{equation}
\vert \psi(t) \rangle =U^*_{\alpha j}e^{-\iota E_j
t/\hbar}U_{\beta j}\vert\nu_\beta (0)\rangle
\label{eq3}
\end{equation}
which in general involve all the three flavors.  In neutrino
oscillation experiments, the propagation time $t$ is not
directly observed, but is instead replaced by the known
distance $L$ between the source and the detector, leading to
the substitution of $t$ by $L/c$. In a matrix
representation, we can express Eqn.~\ref{eq3} as:
\begin{equation}
\vert \psi (t) \rangle=U_{\text{PMNS}} M(t)
U^\dagger_{\text{PMNS}}\vert\nu_\beta (0)\rangle
 \label{eq4}
\end{equation}
with $\hbar=c=1$,
\begin{equation*}
M(t)=\begin{pmatrix}
1&0&0\\
0&e^{-\iota \frac{\Delta m^2_{21}L}{2 E}}&0\\
0&0&e^{-\iota \frac{\Delta m^2_{31}L}{2 E}}\\
\end{pmatrix}
\end{equation*}
where $\Delta m^2_{ij}=m^2_i-m^2_j$ 
denotes neutrino mass squared difference and 
\begin{equation*}
\frac{\Delta m^2_{ij}L}{2 E}=2\times
1.27\times \Delta
m^2_{ij}{(\text{eV}^2)}\frac{L\text{(km)}}{E\text{(GeV)}}
\end{equation*}
The probability of detection of neutrino flavor $\nu_\beta$ from $\nu_\alpha$
is obtained as
\begin{equation}
P_{\nu_\alpha\rightarrow\nu_\beta}=|\langle{\nu_\beta}\vert
{\psi(t)}\rangle|^2= \sum_{i,j}U_{\alpha i}^*U_{\beta
i}U_{\alpha j}U_{\beta j}^* e^{-\iota \frac{\Delta
m_{ij}^2L}{2E}}\label{eq5}
\end{equation} 
The above expression indicates that the phases of neutrino oscillations are
determined by the experiment-dependent quantities $L$ and $E$. For large $L/E$,
neutrino oscillations are experimentally accessible even for tiny masses of
neutrinos.  The PMNS matrix governing neutrino oscillations is also
characterized by a complex phase $\delta$, that leads to the dependence of the
oscillations on $\delta$, enabling the observation of CP symmetry violation
as the neutrino and the anti-neutrino oscillate with different
probabilities. 
\subsection{Neutrino Oscillations in Matter}
\label{sec2b}
The modification in the mixing of neutrinos during propagation in matter
was first observed by Wolfenstein~\cite{wolfenstein-prd-1978} and governed by a
modified Hamiltonian which
contains two terms~\cite{ioannisian-plb-2018,nguyen-prd-2023}:
the vacuum Hamiltonian in 
the flavor basis $H_0$, and the potential term arising due to
interaction of neutrinos in matter $V_m$~\cite{gonzalezgarcia-pr-2008}. The
modified Hamiltonian is defined as: 
\begin{equation}
\footnotesize
H=\underbrace{\frac{1}{2E}U_{\text{PMNS}} \begin{pmatrix}
0&0&0\\
0&\Delta m^2_{21}&0\\
0&0&\Delta m^2_{31}\\
\end{pmatrix} U^\dagger_{\text{PMNS}}}_{\substack{H_0}}+\underbrace{\begin{pmatrix}
V&0&0\\
0&0&0\\
0&0&0\\
\end{pmatrix}}_{\substack{V_m}}\label{eq7}
\end{equation}
where $V=\sqrt{2}G_F N_e$ represents the
Wolfenstein matter potential~\cite{wolfenstein-prd-1978}, which is induced by
coherent forward scattering of neutrinos on electrons in matter; $G_F$ is the
Fermi coupling constant, and $N_e$ is the number density of electrons in 
matter. The matter interaction term (potential term) can be regarded as a
perturbation to the vacuum Hamiltonian.

Due to the presence of this interaction term, the mass eigenstates governing
neutrino propagation in matter differ from the mass eigenstates
$\vert\nu_j\rangle$ observed in  vacuum. Instead, they are matter eigenstates
$\vert\tilde{\nu}_j\rangle$, which connects to the flavor eigenstates through
the matter mixing matrix $\tilde{U}_{\text{PMNS}}$. This matrix encodes the
modified mixing angles and diagonalizes the Hamiltonian while preserving a
similar form to the vacuum case as: 
\begin{equation}
H=\frac{1}{2E}\tilde{U}_{\text{PMNS}} \begin{pmatrix}
0&0&0\\
0&\Delta \tilde{m}^2_{21}&0\\
0&0&\Delta \tilde{m}^2_{31}\\
\end{pmatrix} \tilde{U}^\dagger_{\text{PMNS}}\label{eq8}
\end{equation}
where $\tilde{m}^2_{21}$ and $\tilde{m}^2_{31}$ are mass squared differences
for matter interactions~\cite{ioannisian-plb-2018}.
\begin{equation}
\begin{split}
\Delta\tilde{m}^2_{21}=&\Delta{m}^2_{21}\sqrt{(\cos 2\theta_{12}-\epsilon_1)^2+(\sin 2\theta_{12}\cos 2\phi_{13})^2}\\
\Delta\tilde{m}^2_{31}=&\frac{3}{4}\Delta{m}^2_{ee} \sqrt{(\cos 2\theta_{13}-\epsilon_2)^2+\sin^2 2\theta_{13}} \\
&+\frac{1}{4}(\Delta{m}^2_{ee}+a)+\frac{1}{2}(\Delta\tilde{m}^2_{21}-\Delta{m}^2_{21}\cos 2\theta_{12})
\end{split}
\label{eq9}
\end{equation}
with
\begin{equation*}
\begin{split}
\epsilon_1&=\frac{a}{\Delta{m}^2_{21}} \cos^2 (\phi_{13}+\theta_{13})+\Delta{m}^2_{ee}\frac{\sin^2 \phi_{13}}{\Delta{m}^2_{21}}\\
\epsilon_2&=a/\Delta{m}^2_{ee}\\
\Delta{m}^2_{ee}&=\Delta{m}^2_{31} \cos^2\theta_{12}+\Delta{m}^2_{32} \sin^2\theta_{12}\\
a&=2EV\approx \frac{V (\sim 10^{-4} \rho_m)}{ (\text{eV})} \frac{E}{(\text{GeV})}\, \text{eV}^2
\end{split}
\end{equation*}
where $\rho_m$(g/cm$^3$) is the density of matter and $\Delta{m}^2_{ee}$ is the
effective mass squared difference for electron flavor~\cite{parke-prd-2016}.
The matter mixing matrix $\tilde{U}_{\text{PMNS}}\equiv
\tilde{U}_{\text{PMNS}}(\tilde{\theta}_{12},\tilde{\theta}_{23},\tilde{\theta}_{13},\tilde{\delta})$
is characterized by modified $\tilde{\theta}_{ij}$ and $\tilde{\delta}$ as
defined in~\citep{ioannisian-plb-2018}.
\begin{equation}
\begin{split}
\cos 2 \tilde{\theta}_{13} = &
\frac{\cos 2 \theta_{13}-\epsilon_2}{ \sqrt{(\cos 2 \theta_{13}-\epsilon_2)^2+\sin^2 2 \theta_{13}}}\\
\cos 2 \tilde{\theta}_{12} = &
\frac{\cos 2 \theta_{12}-\epsilon_1}{ \sqrt{(\cos 2 \theta_{12}-\epsilon_1)^2+(\sin 2 \theta_{12}\sin 2 \phi_{13})^2}}\\
\tilde{\theta}_{23}\equiv&\, {\theta}_{23}\\
\tilde{\delta}\equiv&\,\delta
\end{split}\label{eq10}
\end{equation}
\subsection{Simulating Three-Flavor Neutrino Oscillations in a Two-Qubit
Subspace}
\label{sec2c}
To simulate three-flavor neutrino oscillations on 
a two-qubit system, we
mapped the three neutrino flavor eigenstates
$\{\vert\nu_e\rangle,\vert\nu_\mu\rangle,\vert\nu_\tau\rangle\}$ to
the orthogonal states of a two-qubit system. The remaining basis is assigned to
represent the fourth neutrino flavor, i.e., the light
sterile neutrino~\cite{gariazzo_jpg_2016} $\vert\nu_\chi\rangle$ as:
\begin{equation}
\vert\nu_e\rangle=\begin{pmatrix}
1\\
0\\
0\\
0
\end{pmatrix}\hspace{0.3cm}
\vert\nu_\mu\rangle=\begin{pmatrix}
0\\
1\\
0\\
0
\end{pmatrix}\hspace{0.3cm}
\vert\nu_\tau\rangle=\begin{pmatrix}
0\\
0\\
1\\
0
\end{pmatrix}\hspace{0.3cm}
\vert\nu_\chi\rangle=\begin{pmatrix}
0\\
0\\
0\\
1
\end{pmatrix}
\end{equation}
A sterile neutrino is a neutral fermion that does not participate in weak
interactions,  and in our work, we have considered sterile neutrinos to be
physically decoupled. Thus, the PMNS matrix becomes a $4\times 4$ unitary matrix
connecting flavor eigenstates
$\{\vert\nu_e\rangle,\vert\nu_\mu\rangle,\vert\nu_\tau\rangle,\vert\nu_\chi\rangle\}$
to mass eigenstates
$\{\vert\nu_1\rangle,\vert\nu_2\rangle,\vert\nu_3\rangle,\vert\nu_4\rangle\}$
and Eqn.~\ref{eq2} gets modified to:
\begin{equation}
U_{\text{PMNS}}=\begin{pmatrix}
 U_{e1}&  U_{e2}&U_{e3}&0  \\
 U_{\mu 1}&  U_{\mu 2}&U_{\mu 3}&0  \\
 U_{\tau 1}&  U_{\tau 2}&U_{\tau 3} &0 \\
 0&0&0&1
\end{pmatrix}
\end{equation}
and $M(t)$ is modified to:
\begin{equation}
M(t)=\begin{pmatrix}
1&0&0&0\\
0&e^{-\iota \frac{\Delta m^2_{21}L}{2 E}}&0&0\\
0&0&e^{-\iota \frac{\Delta m^2_{31}L}{2 E}}&0\\
0&0&0&e^{-\iota \Phi_{\text{ab}}}
\end{pmatrix}
\end{equation}
The phase $\Phi_{\text{ab}}$ can be chosen conveniently without impacting
observable quantities, offering phase freedom in the $\nu_\chi$ basis. 
It enables
the simulation of the time evolution operator $M(t)$ by employing only single
qubit $z$-rotations.
\begin{figure}[t!]
\centering
{\includegraphics[scale=0.9]{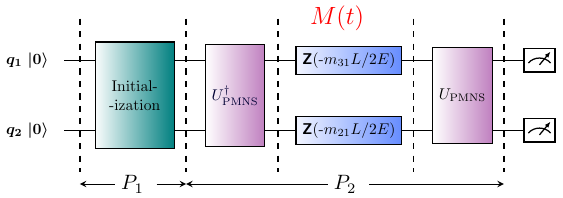}}
\caption{Quantum circuit using the PMNS framework for three-flavor
 neutrino oscillations, starting with flavor state initialization on 
 a two-qubit system. }
\label{fig1}
\end{figure}
 \begin{figure*}[t!]
\centering
{\includegraphics[scale=1]{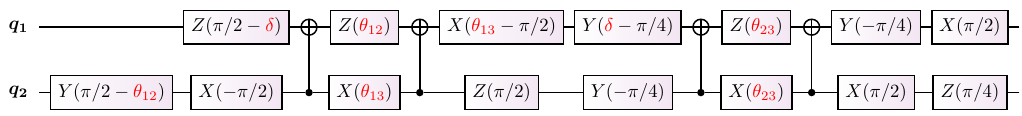}}
\caption{Optimized quantum circuit designed to implement the
PMNS matrix on a two-qubit system, encoding all four parameters
$\{\theta_{12}, \theta_{23}, \theta_{13}, \delta\}$. }
\label{fig2}
\end{figure*}

Figure~\ref{fig1} illustrates the quantum circuit designed for simulating
three-flavor neutrino oscillations. Initially, the system is initialized in one
of the flavor states
$\{\vert\nu_e\rangle,\vert\nu_\mu\rangle,\vert\nu_\tau\rangle\}$. Subsequently,
the PMNS matrix is employed to transform to the mass basis, as outlined in
Eqn.~\ref{eq1}. The state then undergoes evolution according to Eqn.~\ref{eq3},
followed by another transformation from the mass basis back to the flavor basis
using the PMNS matrix. 

A new circuit using controlled rotations was designed to encode the PMNS matrix
in three and higher dimensions and to incorporate the CP violating
phase~\cite{molewski-prd-2022}.  According to this novel circuit, a two-qubit
system necessitates three controlled rotations and two CNOT gates to implement
the PMNS matrix encoding four parameters (three mixing angles and one complex
phase). While this circuit is efficient and provides physical insights into all
four parameters, further optimization is required for implementation on
experimental platforms such as NMR quantum processors, where increased gate
complexity introduces decoherence effects, significantly impacting fidelity.
Therefore, we introduced an optimized quantum circuit utilizing four CNOT gates
combined with single-qubit rotations to implement the PMNS matrix on a
two-qubit system (Figure~\ref{fig2}).  For simulations of neutrino oscillations
in the presence of matter, the same quantum circuit is used (Figure~\ref{fig1})
with modified parameters (Eqn.~\ref{eq10})
$(\tilde{\theta}_{12},\tilde{\theta}_{13},\tilde{\theta}_{23},\tilde{\delta},\tilde{m}_{21},\tilde{m}_{31})$.
\section{Experimental implementation on 
an NMR Quantum Processor}
\label{sec3}
\subsection{Initial State Preparation}
\label{sec3a}
\begin{figure}[t!]
\centering
{\includegraphics[scale=0.9]{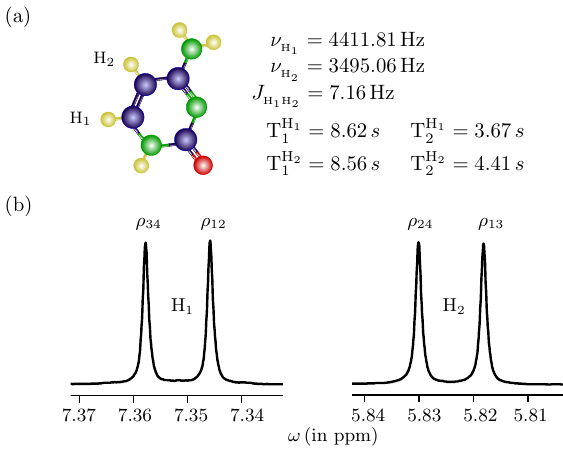}} 
\caption{(a)
Molecular structure of cytosine molecule, along with its Hamiltonian
parameters, including Larmor frequencies, scalar J-coupling and
relaxation times T$_1$ and T$_2$ for Hydrogen nuclei H$_1$ and H$_2$.
Purple, green, red, and yellow balls represent the carbon, nitrogen,
oxygen, and hydrogen nuclei, respectively. (b) NMR spectra (real:
absorption mode) corresponding to the hydrogen nuclei of cytosine
molecule obtained by applying a $\pi/2$ readout pulse on 
the thermal
equilibrium state. The spectral lines of each qubit, which represent
the elements $\rho_{12}$, $\rho_{34}$, $\rho_{13}$, and $\rho_{24}$ of
the density matrix $\rho$, are delineated in the spectrum.}
\label{fig3}
\end{figure}

Neutrino oscillations were simulated using 
a two-qubit NMR quantum information
processor.  All the experiments were performed on 
a Bruker DRX Avance-III 600 MHz
NMR spectrometer equipped with a standard 5 mm QXI probe at room temperature
$\approx 300$K, with  the power level set to be 18.14 W in all the
experiments. The qubits are encoded by 
the two protons of cytosine molecule
dissolved in deuterated solvent D$_2$O. The molecular structure and  molecular
parameters 
are listed in Figure~\ref{fig3}. 
The Hamiltonian of a two spin-1/2 system, under weak coupling
approximation, in the rotating frame rotating with a frequency $\omega_{rf}$ is
expressed as~\cite{oliveira-book-2007}
\begin{equation}
\mathcal{H}=-\sum_{i=1}^{2} (\omega_{i}-\omega_{rf})  I_z^i+2\pi\sum_{i, j=1, i>j}^{2} J_{i j} I_z^i  I_z^j\label{eq5}
\end{equation}
where $\omega_i=2\pi\nu_i$ is the Larmor frequency and $I_z^i$ is the $z$
component of spin angular momentum of $i$th nuclei.  We used the spatial
averaging technique~\cite{oliveira-book-2007} to initialize the system in a
pseudo-pure state (PPS) $\vert 00\rangle$ with the corresponding density
operator 
\begin{equation}
\rho_{00}=\frac{1-\eta}{2^2}\mathbb{I}_{4\times 4}+\eta\vert 00\rangle\langle
00 \vert
\end{equation}
where $\eta\sim 10^{-5}$ is the thermal polarization 
and $\mathbb{I}_{4\times 4}$
is the Identity operator. We employed the Uhlmann-Jozsa fidelity measure~\cite{uhlmann-rmp-1976,jozsa-jmo-1994} which calculates the projection between
theoretically expected density matrix $\rho_{\text{th}}$ and experimentally
reconstructed density matrix $\rho_{\text{exp}}$, to measure the fidelity of
an experimentally reconstructed density matrix as:
\begin{equation}
\mathcal{F}=\left[\text{Tr}\left[\sqrt{\sqrt{\rho_{\text{th}}}\rho_{\text{exp}}\sqrt{\rho_{\text{th}}}}\right]\right]^2
\end{equation} 
In this work, we used the constrained convex optimization based reduced state
tomography method~\cite{gaikwad-qip-2021} to reconstruct a physically valid
density matrix. We used the NMR pulse sequence given in
Reference~\cite{harpreet-pra-2017} to prepare the PPS from the thermal
equilibrium state. The PPS is then experimentally tomographed with fidelity
$\mathcal{F}=0.997\pm0.0012$ using a set of tomographic pulses
$\{II,IX,IY,XX\}$, where, $X(Y)$ denotes a spin-selective $\pi/2$ rotation
along $x(y)$ axis and $I$ is the identity operation (do nothing). All
single-qubit selective excitations with duration ranging from approximately 350
to 700 $\mu$s and an average fidelity $\geq 0.999$ were optimized using the
Gradient Ascent Pulse Engineering (GRAPE)
algorithm~\cite{khaneja-jmr-2005,tosner-jmr-2009,gsingh-qip-2023}
and were crafted to be robust against rf inhomogeneity.
\subsection{Experimental Simulation of Neutrino Oscillations}
\label{sec3b}
\begin{figure*}[t!]
\centering
{\includegraphics[scale=1]{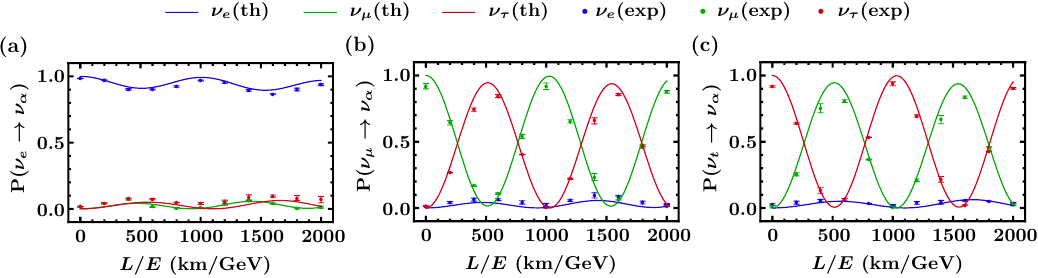}}
\caption{The three-flavor neutrino oscillation probability as a function of
standard scale $L (\text{km})/E (\text{GeV})$: solid lines represent
the numerical simulations and dots with error bars corresponds to the
experimental simulations on NMR for various initial states: (a)
electron neutrino, (b) muon neutrino, and (c) tauon neutrino. The
colors blue, green, and red represent the final states
$\nu_\alpha: \nu_e, \nu_\mu$ and $\nu_\tau$, respectively.}
\label{fig4}
\end{figure*}
\begin{figure*}[t!]
\centering
{\includegraphics[scale=1]{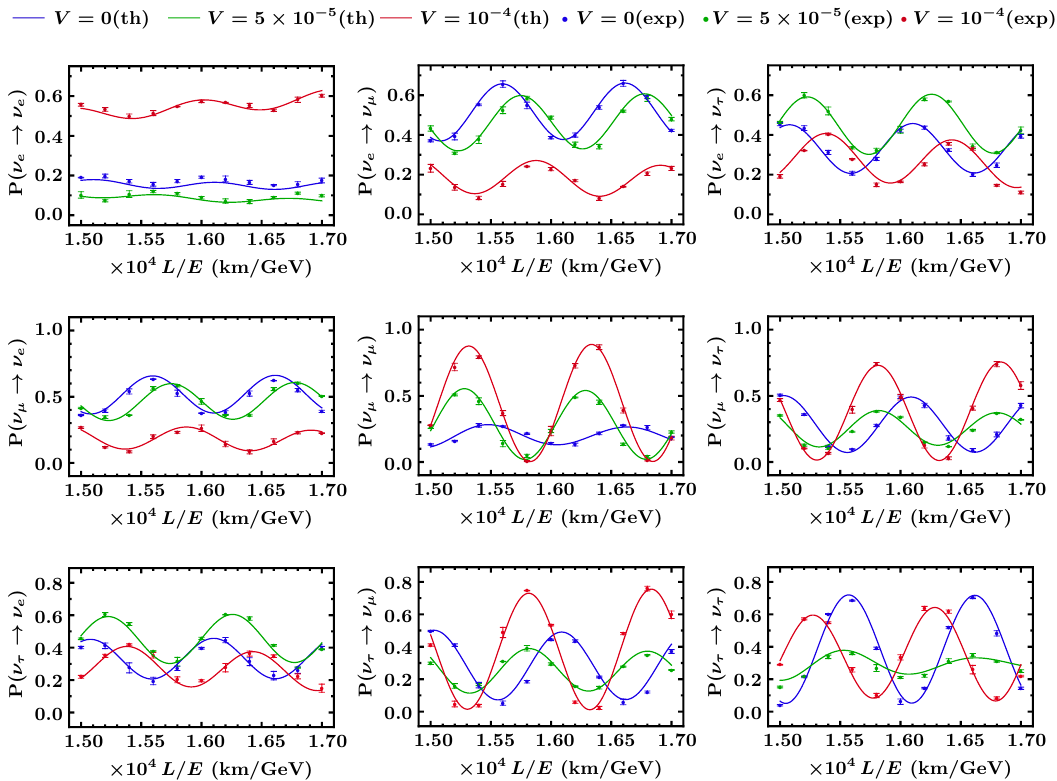}}
\caption{Comparison of neutrino oscillations probability of as a function of
standard scale $L (\text{km})/E (\text{GeV})$ with energy $E=0.5$ GeV
in presence of matter with potential $V=0$ eV (blue), $V=5\times
10^{-5}$ eV (green) and $V=10^{-4}$ eV (red). In this illustration,
each row corresponds to the initial state, and each column corresponds
to the final state of the neutrino $\nu_e,\nu_\mu$ and $\nu_\tau$
respectively. The solid curve depicts the numerically simulated
results, while the dots represent the experimentally simulated
results.}
\label{fig5}
\end{figure*}
The quantum circuit depicted in Figure~\ref{fig1} is
implemented on NMR quantum processor utilizing
GRAPE-optimized radio-frequency pulses. For experimental
implementation $P_1$ pulse with duration $\sim$ 650-700
$\mu$s is used for state initialization and a single
GRAPE-optimized pulse $P_2$ was crafted for entire unitary
operation $U_{\text{PMNS}}M(t)U^\dagger_{\text{PMNS}}$ with
a pulse duration ranging from 5000 to 18000 $\mu$s for
various $L/E$ value in case of vacuum and matter
interaction. After implementing the entire unitary operation
(Eqn.~\ref{eq4}), the resulting density matrix can be
expressed generally as
\begin{equation}
\rho=\begin{pmatrix}
\rho_{11}&\rho_{12}&\rho_{13}&\rho_{14}\\
\rho_{12}^*&\rho_{22}&\rho_{23}&\rho_{24}\\
\rho_{13}^*&\rho_{23}^*&\rho_{33}&\rho_{34}\\
\rho_{14}^*&\rho_{24}^*&\rho_{34}^*&\rho_{44}\\
\end{pmatrix}
\end{equation}
Following this, the diagonal entries of $\rho$ corresponds
to the oscillation probabilities as $P(\nu_\alpha\rightarrow
\nu_e)=\rho_{11},P(\nu_\alpha\rightarrow \nu_\mu)=\rho_{22}$
and $P(\nu_\alpha\rightarrow \nu_\tau)=\rho_{33}$.

The NMR spectra of cytosine molecule (Figure~\ref{fig3})
include NMR signal for protons H$_1$ and H$_2$, each
consists of two spectral peaks corresponding to the
transitions associated with specific density matrix
elements, known as readout elements: $\rho_{12},\rho_{34}$
from spectrum corresponding to H$_1$ and
$\rho_{13},\rho_{24}$ from spectrum corresponding to H$_2$.
The diagonal elements of the density matrix, however, are
not directly observable. To determine them, one 
 has to either reconstruct the full density matrix or the
expectation values $\langle I^1_z\rangle,\langle
I^2_z\rangle$ and $\langle I^1_zI^2_z \rangle$ operators
should be evaluated. As discussed in the previous section,
four tomographic pulses are required to reconstruct the full
experimental density matrix. Additionally, to directly
calculate the expectation values, we have to map to the
local z-magnetization of one of the qubits followed by the
implementation of the corresponding unitary operations to
obtain each expectation value~\cite{amandeep-pra-2016}.

Rather than employing the aforementioned methods, we have
embraced an alternative approach. Through the acquisition of
NMR signals from H$_1$ and H$_2$ following the
implementation of only two acquisition pulses $\{YI,IY\}$,
we can obtain the diagonal entries $\rho_{11},\rho_{22}$ and
$\rho_{33}$ for each $L/E$ values. After implementing
acquisition pulse $YI$ and $IY$, the state is mapped into
another state as $\rho^1=YI.\rho.YI^\dagger$ and
$\rho^2=IY.\rho.IY^\dagger$ respectively. The line
intensities in the absorption mode of NMR spectra (real part
of readout element) of these mapped states contain
information about the diagonal entries of $\rho$ and it can
be mathematically expressed as
\begin{equation}
\begin{split}
\rho_{11}=&(1+2 (\rho^1_{13}+\rho^1_{24})+4 \rho^2_{12})/4\\
\rho_{22}=&(1+2 (\rho^1_{13}+\rho^1_{24})-4 \rho^2_{12})/4\\
\rho_{33}=&(1-2 (\rho^1_{13}+\rho^1_{24})+4 \rho^2_{34})/4\\
\end{split}
\end{equation}

where, $\rho^i_{jk}$ is the spectral line intensity of
mapped state $\rho^i$ with respect to the readout element
$jk$.

Figure~\ref{fig4} illustrates the vacuum oscillations in
three flavor neutrino system, without incorporating CP
violating phase $\delta$, focussing on a smaller L/E range,
where oscillations are approximately two flavor dominant.
The plots show the probability of detecting final neutrino
flavor state $\nu_\alpha$ as a function of standard scale $L
(\text{km})/E (\text{GeV})$ starting from three different
initial states: electron neutrino (Figure~\ref{fig4}$(\mathrm{a})$), 
muon neutrino (Figure~\ref{fig4}$(\mathrm{b})$) and tau neutrino
(Figure~\ref{fig4}$(\mathrm{c})$). The final neutrino state
$\nu_\alpha:\,\nu_e,\nu_\mu$ and $\nu_\tau$ are visually
distinguished by the blue, green and red colors
respectively. Solid lines correspond to results from
numerical simulations, while experimental data is
represented by dots with error bars.

In the case of interaction with matter, we treated the
potential term as a perturbation to the vacuum Hamiltonian
and diagonalized the total Hamiltonian to maintain the same
form as in the vacuum scenario. As a result, we implemented
the identical quantum circuit illustrated in
Figure~\ref{fig1}, \ref{fig2} with modified parameters
$(\tilde{\theta}_{12},\tilde{\theta}_{13},\tilde{\theta}_{23},\tilde{m}_{21},\tilde{m}_{31})$
and $\delta=0$ for different value of $V$. We selected three
distinct values for $V=0$, $5\times 10^{-5}$, and $10^{-4}$
eV to examine the impact of the presence of matter
interaction and simulated neutrino oscillations for all
three neutrino states as the initial state. The
experimentally simulated and numerically simulated
oscillation probabilities starting with different initial
states for $\delta=0$ are depicted in Figure~\ref{fig5}. In
this figure, row represents the initial state
$(\nu_e,\nu_\mu, \nu_\tau)$ and column represents the final
state $(\nu_e,\nu_\mu, \nu_\tau)$. For instance, the plot in
the second column of the first row illustrates the
oscillation probability of detecting $\nu_\mu$ starting from
an initial state of $\nu_e$ under different potentials.

 \begin{figure}[h!]
\centering
{\includegraphics[scale=0.9]{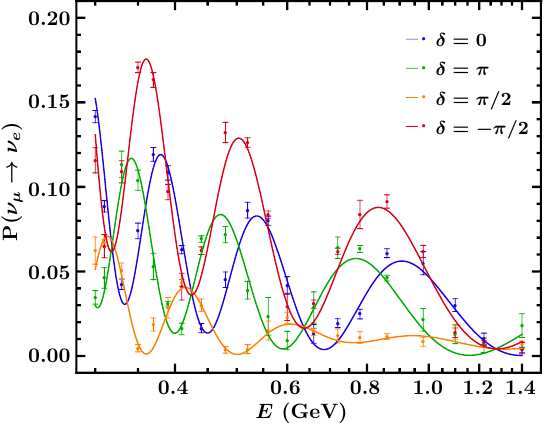}}
\caption{The oscillations probability of $\nu_\mu\rightarrow\nu_e$ as a
 function of $E$ (GeV) across the DUNE baseline length of 1298 km is
 evaluated for different CP-violating phases: $\delta=0$, $\pi/2$,
 $\pi$, and $-\pi/2$, under vacuum conditions ($V=0$ eV). The numerical
 simulations are depicted by solid curves, while the dots with error
 bars represent the experimental data.}
\label{fig6}
\end{figure}

We also showcased the efficacy of the quantum circuit in implementing the PMNS
matrix, incorporating a CP-violating phase $\delta\neq 0$. Figure~\ref{fig6}
illustrates the oscillation probabilities of $\nu_e$ appearance starting with
the initial state of $\nu_\mu$ as a function of energy E (GeV) over a baseline
length of $L=$1298 km (reflecting DUNE experimental configuration). In this
plot, dots with error bars represents the experimental data, while the solid
lines are the numerical results for different values of
$\delta=0,\pi/2,\pi,-\pi/2$. CP violating phase $\delta$ alters the atmospheric
oscillations by introducing both a phase shift and a constant offset, with the
oscillation probability containing terms proportional to $\sin \delta$ and
$\cos \delta$. These terms influence the interference between mass eigenstates
as they propagate, leading to constructive interference for $\delta=-\pi/2$ and
destructive interference in case of $\delta=\pi/2$ (Figure~\ref{fig6}).
Although both $\delta=0$ and $\pi$ correspond to the scenario without CP
violation, since the Hamiltonian is real, the oscillations probabilities differ
in these two cases due to distinct interference. 

 \begin{figure}[h!]
\centering
{\includegraphics[scale=0.75]{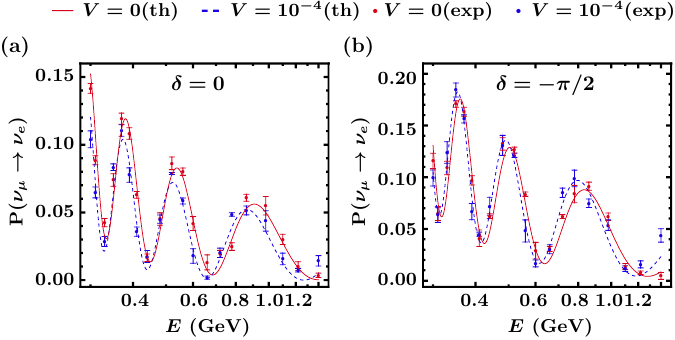}}
\caption{A comparative study of oscillation probabilities of
 $\nu_\mu\rightarrow\nu_e$ as a function of energy ($E$ in GeV) for a
 baseline length of 1298 km (which is the distance between the source
 and the detector in the DUNE experiment). The comparison was made for
 oscillations with potential $V=0$ eV and $V=10^{-4}$ eV, considering
 two different values of the CP-violating phase $\delta=0$ and
 $\delta=-\pi/2$. The results are illustrated by solid red curves for
 $V=0$ eV and dashed green curves for $V=10^{-4}$ eV in numerical
 simulations, while experimental data are represented by dots
 with error bars.}
\label{fig7}
\end{figure}

We then experimentally simulated the oscillations with the
initial state of the muon neutrino $\nu_\mu$ for $V=10^{-4}$
eV at CP-violating phase $\delta=0$ and $-\pi/2$, and
compared the oscillation probabilities of $\nu_e$ appearance
with the vacuum scenario ($V=0$ eV). Figure~\ref{fig7}
illustrates this comparison. In the context of the DUNE
distance, the primary factor influencing oscillation
probabilities is the change in the angle $\theta_{13}$.
However, it is noteworthy that the change in the angle
$\theta_{13}$ is relatively small, which results in small
differences between vacuum and matter oscillations, but
exhibit qualitatively similar
characteristics~\cite{ioannisian-plb-2018}.

All the experimental simulations are in good agreement with
the numerical simulations, effectively demonstrating an
interplay between neutrino mixing parameter and the complex
phase $\delta$ in three-flavor neutrino oscillations.  This
highlights the successful implementation of the PMNS
framework to simulate neutrino oscillations on an NMR
quantum processor.
\section{Conclusion}
\label{sec4}
We successfully demonstrated the quantum simulation of three-flavor neutrino
oscillations using PMNS theory on a subspace of a two-qubit NMR quantum
information processor.  We highlighted the efficacy of the proposed quantum
circuit for implementing the PMNS matrix encoding the parameters, $\theta_{12},
\theta_{13}, \theta_{23}$, and $\delta$. This circuit employs single rotations
and four CNOT gates and is experimentally feasible to implement using
GRAPE-optimized rf pulses, thus reducing circuit complexity and errors.
Furthermore, we presented an alternative approach (specific to the NMR
platform) to directly obtain probabilities from NMR spectra, obviating the need
for reconstructing the full density matrix by computing expectation values of
$\langle I^1_z\rangle, \langle I^2_z\rangle$, and $\langle I^1_zI^2_z \rangle$
operators. Our experimental results closely align with the theoretical
simulations of PMNS theory, encompassing scenarios with and without
CP-violating phase $\delta$ for both, oscillations in vacuum and interaction
with matter. A comparative analysis was made for the vacuum by incorporating CP
phase $\delta=0,\pi/2,\pi$ and $-\pi/2$.  We showcased the performance of the
optimized quantum circuit by employing it for matter interaction case,
incorporating the CP-violating phase and then comparing it with the vacuum
scenario. 
 
The advantage of quantum simulators beyond their classical counterparts becomes
apparent when addressing complex scenarios, such as collective neutrino
oscillations, and correlations in neutrino-neutrino interactions, which is a
many-body problem with a computational cost that scales exponentially with the
number of neutrinos \cite{bauer-prxq-2023}. Moreover, using quantum
simulations, out-of-equilibrium flavor dynamics can be explored in regimes
inaccessible by classical approaches~\cite{amitrano-prd-2023}.  Quantum
computers have the potential to explore scenarios beyond the Standard Model,
such as the existence of sterile neutrinos to explain experimental anomalies
that cannot be accounted for by the three-flavor oscillation framework and to
probe properties of neutrinos beyond their masses and mixing, or interactions
that deviate from weak interactions~\cite{miranda-njp-2015}.  Our study
explores the potential of quantum simulators such as NMR quantum processors as
a valuable tool to understand high-energy physics phenomena, shedding light on
aspects that present challenges for direct experimental observation.

\begin{acknowledgments}
All experiments were performed on a Bruker DRX Avance-III
600 MHz FT-NMR spectrometer at the NMR Research Facility at
IISER Mohali. G.S.  acknowledges University Grants
Commission (UGC), India for financial support.
\end{acknowledgments}

%

\end{document}